# Open Access –Towards a non-normative and systematic understanding

Niels Taubert[1], Anne Hobert[2], Nick Fraser[3], Najko Jahn[4], Elham Iravani[5]

[1] niels.taubert@uni-bielefeld.de
AG Bibliometrics, Bielefeld University p.o.box 10 01 31, 33501 Bielefeld, Germany

[2] hobert@sub.uni-goettingen.de
SUB University of Göttingen Platz der Göttinger Sieben 1, 37073 Göttingen, Germany

[3] N.Fraser@zbw.eu
Leibniz Information Centre for Economics (ZBW), Düsternbrooker Weg 120, 24105 Kiel, Germany

[4] najko.jahn@sub.uni-goettingen.de
SUB University of Göttingen Platz der Göttinger Sieben 1, 37073 Göttingen, Germany

[5] eiravani@techfak.uni-bielefeld.de
AG Bibliometrics, Bielefeld University p.o.box 10 01 31, 33501 Bielefeld, Germany

## Abstract

The term Open Access not only describes a certain model of scholarly publishing – namely in digital format freely accessible to readers – but often also implies that free availability of research results is desirable, and hence has a normative character. Together with the large variety of presently used definitions of different Open Access types, this normativity hinders a systematic investigation of the development of open availability of scholarly literature. In this paper, we propose a non-normative definition of Open Access and its usage as a neutral, descriptive term in bibliometric studies and research on science. To this end, we first specify what normative figures are commonly associated with the term Open Access and then develop a neutral definition. We further identify distinguishing characteristics of openly accessible literature, called dimensions, and derive a classification scheme into Open Access categories based on these dimensions. Additionally, we present an operationalisation method to assign scientific publications to the respective categories in practice. Here, we describe useful data sources, which can be employed to gather the information needed for the classification of scholarly works according to the presented classification scheme.

## 1. Introduction

Open Access (OA) has become an ambiguous term. Coined in a series of declaration following up the *Budapest Open Access Initiative*, it describes an already existing practice of scientists that evolved during the late 1980s and early 1990s: the use of digital means to freely disseminate research without any restrictions other than those inseparable from gaining access to the internet (BOAI, 2002). The context in which the term originated has impregnated it. OA does not only refer to a publication practice but also implies that free digital access to research results is an objective worth pursuing. The term has thus developed a normative tongue, and the search for a precise definition has become a battlefield between OA proponents, publishers and other stakeholders that each try to push a specific understanding, sometimes also linked to a business model. Concurrent to these discussions surrounding OA definitions, OA has become a major issue in science policy; given that political targets have been established, expressed in OA shares to be reached at a specific point in time (e.g. European Commission 2012, 2019, Abgeordnetenhaus Berlin 2015), the necessity arose to measure and monitor the pick-up of OA and the dynamics of the transformation.[1]

Without doubt, the growing interest in studying OA from an analytical perspective is pleasing, but a problematic aspect remains, in that the results of studies reporting OA shares for the same entities of the research system differ, sometimes to a considerable extent.[2] The reason for that is at least threefold. First, OA shares are calculated based on different citation databases including the Web of Science (WoS), Scopus, Dimensions, or local systems like research databases, repositories or Current Research Information Systems (CRIS) that cover different parts of the publication output of an entity. Second, a number of terms describing the variety of different manifestations of OA were invented and introduced, sometimes in a rather ad-hoc manner and without systematic consideration of how they relate to existing ones. Beginning with the well-established types such as Green and Gold (Suber 2012, Pinfield 2015), a number of other types of OA have subsequently emerged (see Table 1).

To some extent, the inventiveness reflects the diversity of how free electronic access to research articles is put into practice, but it also results from studies on the dispersion of OA on the basis of incomplete metadata.[3] In addition, the definitions of the same classes of OA publications differ.[4] Third, the aforementioned databases do not provide OA information that is sufficient for all types of use or analysis. Therefore, they are often enriched with OA information from secondary sources. The operationalization of the OA types, the sources of OA information, and the way they are being used may vary in this process.

The current situation can therefore be understood as one of missing standards. This paper takes that as a starting point to rethink the term OA, and to develop a classification scheme. In doing so, we do not intend to reinvent the wheel, but to start a discussion about possible standards and limitations in the field of OA analytics. The paper is organized as follows. In a first step, a working definition of OA is developed. The normative background of the term is explored in a second step, with the aim to distinguish it from its empirical content. In a third step, a more systematic approach for the classification of OA is suggested, where several dimensions, with respect to which the most recently emerged OA types can be distinguished, are identified and the different characteristics of these types are described. In a fourth step, different sources of information about the OA status of a publication are briefly reviewed and operationalisation procedures to identify the different types in practice are suggested. Since the field evolves fast, the operationalisation reflects current possibilities but may be subject to rapid change in future.

---

[1] In Europe, Open Access Monitors are already in place in Denmark, Finland, France, Germany, the Netherlands and the United Kingdom. For an overview, see Knowledge Exchange 2017, pp. 10-19.

[2] For example, OA shares for the German research system reported by Wohlgemuth et al. (2017), the Open Access Monitor (https://open-access-monitor.de), and Abediyarand & Mayr (2019) differ, in some cases significantly.

[3] Bronze OA, for example, can be considered as a residual category since it is defined negatively by missing of a license statement (Piwowar et al. 2018) that would allow specifying to what category the publications belongs. Typically, no license statement is associated with Transient OA either, making it a residual category as well.

[4] Gold OA is a good example here. Some authors define the type as OA delivered by journals (e.g. Suber 2012: 6) and the definition would include hybrid and delayed OA. Other authors restrict the type to media that apply an OA business model (e.g. Schwerpunktinitiative 'Digitale Information' 2011) providing immediate access to all publications. This later definition would exclude Hybrid and Moving Wall OA.

*Table 1: OA-Types in the current discussion*

| *OA Type* | *Description* | *Reference* |
|---|---|---|
| Hybrid OA | Subscription-based journals allow authors to make their individual article immediately available online if article processing charges have been paid. | Prosser 2003 |
| Delayed OA/ Moving Wall OA | Publications are freely available online after an embargo period, which is usually between six and 24 months long. | Willinsky 2003 |
| Platinum OA | Gold OA publications are freely available immediately in fully OA journals without any publications fees to be paid. | Wilson 2007 |
| Diamond OA | Different label for Gold OA publications in fully OA journals without any publication fees to be paid. | Fuchs & Sandoval 2013 |
| Gray OA | Gold OA publications in journals not covered by the Directory of Open Access Journals (DOAJ). | Crawford 2016 |
| Bronze OA | Immediate or delayed OA publications on publishers' websites without any license for reuse. | Piwowar 2018 |
| Transient OA | Publications are openly available only for a certain period of time and are then taken offline or placed behind a paywall. | Archambault et al. 2014 |
| Guerilla OA | Publications that can be freely accessed online but infringe copyrights. | Swartz 2008 |
| Black OA | Publications that can be freely accessed online but infringe copyrights. | Björk 2017 |
| Robin Hood OA | Publications that can be freely accessed online but infringe copyrights. | Archambault et al. 2014 |
| Blue OA | Self-archiving policy of publisher allows deposit of postprint version (final draft). | Hubbard 2007 |
| Yellow OA | Self-archiving policy of publisher allows deposit of preprint version. | Hubbard 2007 |
| White OA | Self-archiving is not formally supported by publishers' policy. | Hubbard 2007 |

## 2. OA: a working definition

Following the turn of the millennium, the term OA first appeared, and then rapidly developed in science policy discussions. It was coined by a series of declarations including the *Budapest Open Access Initiative* (2002), the *Bethesda Statement on Open Access Publishing* (2003), and the *Berlin Declaration on Open Access to Knowledge in the Sciences and Humanities* (2003), that emphatically demanded free access to research output. Because of the shared aims and a shared understanding of these statements, the following discussions referred to them as the BBB-definition (e.g. Arbeitsgruppe Open Access 2009: 8, Suber 2012: 7). At the beginning of the text, the BOAI mentions the communication norms of science that are familiar to sociologists of science:[5]

"An old tradition and a new technology have converged to make possible an unprecedented public good. The old tradition is the willingness of scientists and scholars to publish the fruits of their research in scholarly journals without payment, for the sake of inquiry and knowledge. The new

[5] See Robert K. Merton`s description of the scientific ethos and the norm communism (i.e. understanding the common body of knowledge as a result of a collective effort Merton (1973 [1942])).

technology is the internet. The public good they make possible is the world-wide electronic distribution of the peer-reviewed journal literature and completely free and unrestricted access to it by all scientists, scholars, teachers, students, and other curious minds.“ (Budapest Open Access Initiative 2002)

The main point of reference of the BOAI declaration is the communication norm of science as a free exchange of ideas and results, and the absence of monetary motives on the side of the authors. Together with the free architecture of the internet that allows nearly cost-free dissemination of information, the BOAI formulates the vision to achieve free and comprehensive access to scientific publications:

“By ‘open access‘ to this literature, we mean its free availability on the public internet permitting any user to read, download, copy, distribute, print, search, or link to the full texts of these articles, crawl them for indexing, pass them as data to software, or use them for any other lawful purpose, without financial, legal, or technical barriers other than those inseparable from gaining access to the internet itself. The only constraint on reproduction and distribution, and the only role for copyright in this domain, should be to give authors control over the integrity of their work and the right to be properly acknowledged and cited.“ (Budapest Open Access Initiative 2002)

Although a large number of policy developments to foster OA followed the BOAI[6], it has remained to this date an important reference point regarding the definition of the term. Four aspects of the definition had a strong impact on the understanding of OA in the field: First, OA refers to scientific publications only and not to other types of texts like popular, artists’, or journalists’ writings. Second, BOAI has determined the constitutive characteristics of OA: The term focuses on the role of the recipient in the publication process and implies the absence of technical and financial barriers other than access to the internet. A prerequisite is that the publication has an electronic format, implying that it can be copied at any number, and distributed with the digital infrastructure of the internet. Third, OA relates to the full texts[7] of publications and not to abstracts, parts of the text or to metadata of the publications.[8] Fourth, the name OA suggests that the main characteristic is accessibility. Such an understanding of the term would be incomplete since the BOAI also mentions legal aspects going beyond accessibility, for instance, that an OA license should permit comprehensive use.[9] This typically comprises the rights to read, copy, and re-distribute the work in any format, and in some cases additionally the rights to modify, remix, and transform it. Moreover, it usually includes the obligation to give appropriate credit to the originator(s).

When applying this definition to publications that are freely available on the internet, it soon becomes clear that the BOAI formulates ambitious criteria with respect to the fourth point. Often publications are freely available online but do not come with a license or they are accompanied with more than one, possibly contradicting licenses (Akbaritabar & Stahlschmidt 2019: 7). Even though they are freely accessible online and practically can be used by anyone, they would not fall under the term OA as defined by BOAI. As a consequence, a distinction was made in the following discussion between two OA sub-types, referred to as *gratis OA* and *libre OA* (Suber 2004a; 2004b): The broader term *gratis OA* indicates access free of charge and the absence of financial barriers, while *libre OA* refers to online access free of charge plus some additional re-use rights beyond fair-use principles (Suber 2012: 6).

---

6 Declarations in favor of OA can be found on global (e.g. GRC 2013; ICSU 2014; OECD 2015), European (Science Europe n.d.; European Commission 2013) and national level (e.g. BMBF 2016) as well as on the level of federal states (e.g. Ministerium für Wissenschaft, Forschung und Kunst Baden-Württemberg 2014: 14ff.; Abgeordnetenhaus Berlin 2015). In addition, many universities, research institutions and funding councils have established an OA policy that encourages, or requires researchers to publish OA. The directory http://roarmap.eprints.org/ provides a worldwide overview over OA mandates. A typology of different policies can be found in Bailey (2005: 20–25) and Herb (2016).

7 Given that in most cases, *Google Books* does not provide access to the full document but only to a limited number of pages, they are not OA according to this definition.

8 Some of the definitions restrict OA to publications that provide immediate access at the time of publication (e.g. Allianz der deutschen Wissenschaftsorganisationen 2011). We follow a different approach here and include all kinds of publications that are freely accessible online and distinguish different types later.

9 The *Creative-Commons-Licenses* (http://creativecommons.org/) have become by far the most important OA licenses.

At the current state of the development of OA publishing the broader understanding of gratis OA seem to be more suitable to us in the context of bibliometric analysis. But legal aspects should not be fully neglected. Therefore, the definition that is endorsed here also includes the lawfulness of the channels that disseminate the publications. The considerations so far can be summarized in a brief working definition of OA:

*The term Open Access identifies scientific publications that have a digital format and whose full texts can be practically and lawfully accessed and used in some way without any technical and financial restrictions other than those that are connected to accessing the internet.*

## 3. Towards a non-normative understanding of Open Access

The context of the science-policy debate in which the term OA originated does not only affect its exact definition. It has also influenced the wider context of the discourse in which the term is embedded: OA is not a neutral expression in the sense that it can be used as a purely descriptive term but often additionally bears some normative implications. It strives towards specific aims, it is visionary, and makes claims of positive effects of open accessibility which serve as justification. Normativity in this regard does not mean that institutionalized norms of science are described or referred to, like in the BOAI declaration cited above. Instead, the term OA itself tends to be prescriptive since it is accompanied by normative expectations on how science should publish results. In the following, the normative background of the term will be reconstructed. The purpose of this exercise is neither a critique of ideology nor an examination of whether or not the normative arguments are coherent. Instead, the revelation of the normative implications intends to separate them from the descriptive content that was defined in the first step.

When different declarations and policy papers are compared, one frequently comes across four different strands of argument that legitimise OA. They can be labeled as endogenous-utilitarian, exogenous-utilitarian, reciprocity and egalitarian.

*Endogenous-utilitarian argument*: The first and most common type of legitimisation of OA is called *endogenous* as it refers to processes within science and *utilitarian* as OA is legitimized by its usefulness for science. Its persuasiveness results from the additional value of free accessibility and the possibility of unrestricted use of publications compared to the status quo. Following this argument, OA induces some effects that are considered positive because they support the aim of science, namely the increase of knowledge. A number of such effects are mentioned for example in the introductory section of the Action Plan towards OA to Publications, endorsed by the *Global Research Council*:

„[OA] leads to better science e.g. by increasing access to knowledge, by improving the pace and efficiency of research, by enabling computation on research information, and by offering opportunities to foster collaboration and exchange globally." (GRC 2013: 1)[10]

[10] See also the Budapest Open Access Initiative, which includes the factor of acceleration („accelerate research", Budapest Open Access Initiative 2002; 2012).

Besides improved accessibility, and the acceleration and better efficiency of the dissemination of results, an increased reach of publications, and the possibility to reuse their contents within the scientific community is highlighted (Bethesda Statement on OA Publishing 2003). In addition, it is argued that OA helps to avoid unwanted duplication of research (European Commission 2015: 4). The position can be considered as normative for two reasons: First, the positive effects are claims only that usually come mostly without any empirical evidence.[11] Second, the perspective is one-sided since it highlights positive effects but neglects possible unwanted or problematic effects.

*Exogenous-utilitarian argument*: The second type of argument is similar to the first one since OA is promoted by highlighting utility but this time not within science but for the society at large. For this reason, it is called exogenous. An example can be drawn from the *Bethesda Statement on Open Access Publishing* (2003), which states that the mission of science „to maximize public benefit from scientific knowledge" is only half completed "if the work is not made as widely available and as useful to society as possible". Following this argument, the conditions for transfer and application of scientific knowledge are improved by OA in general. Given that these processes are multifaceted, the argument is sketchy here as it mentions only some areas of society that shall benefit from Open Access but does not explain how and under what circumstances.

"Furthermore, increased access to knowledge provides societal benefits to many who rely on research results, be it in patient care, be it in politics and decision making, be it in entrepreneurship or industry, be it in journalism or society at large: there is an enormous need for research information outside universities and research institutes which can be served best by openly accessible research information." (GRC 2013: 1)

Like for the first argument, usually no empirical evidence is given for the extensive claim, and no possible negative or unwanted effects are mentioned.

*Norm of reciprocity*: The third argument differs from the previous ones since it refers to the norm of reciprocity. Free access to research and publications within science and for society at large is justified by the source of the funding. Publicly-funded grants are not understood as a one-way support for science but constitute a claim.[12] The persuasiveness of the argument results from a common conception about equity that, where support is given, a claim of service in return is established. This argument surfaces whenever publicly funded research is referred to as a *public or common good*, like, for example, in OA policy documents of the European Commission:

"The European Commission's vision is that information already paid for by the public purse should not be paid for again each time it is accessed or used, and that it should benefit European companies and citizens to the full. This means making publicly-funded scientific information available online, at no extra cost, to European researchers, innovative industries, and citizens, while ensuring long-term preservation." (European Commission 2015: 4)

*Egalitarian argument*: The fourth and final argument is egalitarian in character and refers to an ideal of humankind that can be achieved with OA. An explicit expression of this vision can be found in the *Budapest Open Access Initiative*:

"Removing access barriers to this literature will […] share the learning of the rich with the poor and the poor with the rich, make this literature as useful as it can be, and lay the foundation for uniting

---

11 For example, there is an ongoing debate about possible citation advantages of OA publications (and loss of impact of non-OA publications). For an overview see: http://opcit.eprints.org/oacitation-biblio.html.

12 This argument is also being used with a stronger emphasis to economic terms and ideas. For example, the Global Research Council uses the term of a "return on investments" GRC (2013: 1) of public funds.

humanity in a common intellectual conversation and quest for knowledge." (Budapest Open Access Initiative 2002)

The quotation brings together an assumption about the effect of OA and a normative vision. The assumption about OA is that the removal of barriers to access is an effective means to reduce inequality and to create the opportunity for mutual learning. However, the normative vision about uniting humanity goes far beyond these merely technical effects and the opportunity of learning from each other. It postulates the idea about the nature of humankind to strive towards knowledge. Given such orientation and given that knowledge is freely available, social gaps – like for example between poor and rich – can be bridged by a uniting principle, namely an intellectual discourse in which everyone can participate.

The reconstruction of the four arguments, which are embedded in the term OA, has been undertaken to distinguish the normative ideas that are associated with it from the descriptive components of the definition as developed in the first section. S uch a step is necessary in order to transfer a term with origins in science policy with the goal to use it to denote a specific type of publishing and to empirically study it.

## 4. Classification of Open Access types

OA covers a variety of phenomena that share common characteristics like digital formatting, free and lawful accessibility and practical usability. However, they differ regarding how access is provided, who is responsible for providing access, which version of the publication is available and at what point in time. The empirical variety is reflected by a large number of different OA types. In the past, when empirical cases did not fall under one of the established types, or when a specific property of a publication was not well reflected by a specific type, often another one was introduced, as described in the first section. On occasion, this happened without any systematic consideration as to how the newly introduced types related to existing ones, leading to an incomplete and fragmented typology. In this section, we suggest a method to classify the most important OA types following a two-step approach. First, several key dimensions are identified that reflect basic empirical characteristics of OA publications. Second, for each dimension we construct a small definite number of OA classes that are complete and mutually exclusive. This strategy aims to prevent the introduction of further OA types in an ad hoc manner.

*Dimension 1 – Location*: The most important distinction is that between Gold and Green OA, and these two types were originally introduced by the BOAI, albeit without using the labels 'Green' and 'Gold' (BOAI 2002). According to this definition, Gold refers to access provided by an OA journal, while Green identifies journal articles that are deposited by scholars in an electronic archive. Given that this definition restricts OA to journal publications only and also confuses the actor that provides access with the place where a publication can be accessed, a slight shift in the definition is needed. We therefore define the first dimension as the *location where access to research results is provided*. Within this dimension, one can distinguish between the formal communication channel (e.g. a journal, conference proceedings, a book or anthology) and all other locations that provide unrestricted access (e.g. repositories, publishers' websites, aggregation portals/harvesting engines). The class Gold *OA* refers to all publications that can freely be accessed via the formal communication channel, while *Green OA* refers to publications that can be accessed lawfully at other places.[13] This distinction between Green OA and Gold OA as fundamental types is common in the literature (e.g. Pinfield 2015: 610, Suber 2012: 53) and in science policy (e.g. European Commission 2015: 2, OECD 2015: 37, Allianz der Deutschen Wissenschaftsorganisationen 2009: 94f.)[14].

[13] Given that the platform SciHub by design offers illegal access to publications and a large number of publications on academic networks like ResearchGate and Academia infringe copyright, content on these platforms is not considered Green OA.

*Dimension 2 – Time*: The second dimension applies to the two basic OA classes that were distinguished above (Gold OA and Green OA), but in different ways. For both, the reference is the point in time when research has been published in the formal communication channel. In the case of Gold OA, we distinguish between *Immediate OA* at the time of publication, and *Delayed OA* or *Moving Wall OA* , where free access is provided at a later point in time after an embargo period.[15] In the case of Green OA, the relevant distinction in the time dimension is that between *preprints* that have been deposited *before* they have been published in the formal communication channel and *postprints* that are deposited at a later point in time.[16]

The following two dimensions each apply to one of the two basic classes, Gold OA or Green OA, only.

*Dimension 3 – Optionality*: The dimension of optionality applies to Gold OA only, since Green OA can always be chosen (if lawfully possible), that is, it is always optional. If all publications in a formal communication channel are openly accessible, there is no choice and OA is the non-optional default. We call this type *Full OA*. If OA is an option that can actively been chosen over a default restricted access model, we call it *Hybrid OA*. (Laakso & Björk 2016: 920).

*Dimension 4 – User group*: The addressed user group can be specified along different criteria. In the case of Green OA, depositing manuscripts on a repository can be restricted to members of a specific organization. This is usually a university or a research institute. The resulting Green subtype is called *Institutional Green OA*. On the other hand, the targeted user group can also be a community of scholars, usually researchers in the same field. This type is called *Disciplinary Green OA*. Please note that this distinction technically does not meet the requirement of completeness since there might be other types of repositories that are different from the two mentioned there. Nevertheless, from a practical point of view these two types cover the most important specifications of user groups of repositories within science.[17]

## 5. Open Access Classification Scheme

The four dimensions of OA described above are collated in the following classification scheme. For better readability, the classification scheme is divided into two tables, where the distinction between tables is made according to the first dimension: the first table displays the subdivision of Gold OA and the second table shows the more detailed partitioning of Green OA.

---

[14] Note that there also is a more narrow definition of Gold OA restricting the type to articles in fully OA journals only.

[15] See Laakso & Björk (2012) and Willinsky (2005).

[16] Following the readers' perspective, most likely the distinction between pre- and postprints with respect to the peer review process is important: preprints do not necessarily have to be subjected to peer review while postprints typically underwent the quality control procedure of the formal communication channel (see Swan 2005). Sometimes, other differences are highlighted in the discussion, for example the version that has been made available (manuscript version, accepted final author version, published version) (OECD 2015: 39).

[17] OpenDOAR the global directory of repositories (http://v2.sherpa.ac.uk/opendoar/) lists 3,663 institutional and 343 disciplinary repositories. 128 repositories aggregate content from other sources and 105 repositories are classified as 'governmental'. The one remaining registered repository is classified as 'undetermined'.

*Table 2: Classification scheme for Gold OA publications.*

| **Dimension** | **Specification** | | |
|---|---|---|---|
| Location providing OA | Formal communication channel *Gold OA* | | |
| Time of OA | Immediate OA at the time of publication | | OA after an embargo period |
| Optionality of OA | Non-optional *Full OA* | Optional *Hybrid OA* | *Delayed OA/Moving Wall OA* |

*Table 2: Classification scheme for Green OA publications.*

| **Dimension** | **Specification** | | | |
|---|---|---|---|---|
| Location providing OA | locations other than the formal communication channel *Green OA* | | | |
| Time of OA | Deposition before publication in formal channel / before peer review process *Preprint* | | Deposition after publication in formal channel / after peer review process *Postprint* | |
| Addressed user group of location providing OA | Institutional repository | Disciplinary repository | Institutional repository | Disciplinary repository |
| | *Institutional Green OA of preprint* | *Disciplinary Green OA of preprint* | *Institutional Green OA of postprint* | *Disciplinary Green OA of postprint* |

## 6. Operationalisation of Open Access classes

In this section, we will focus on journal articles, reviews and conference proceedings only. This focus is chosen because of the practical requirements of the OAUNI project[18] that is based upon on the Web of Science and aims to describe the OA footprints in the publication output of German universities. In this database, the aforementioned document types are most dominant.

In the previous section, we proposed a scheme to identify different classes of openly accessible scholarly output, based on a non-normative understanding of OA. However, putting this classification into practice, that is, assigning a given version of a publication to the correct category is a non-trivial task. The main reason for this is a lack of information and standardization. In bibliometric databases some records contain comprehensive metadata specifying, in particular, an identifier (e.g. DOI), the place of publication, the publication and open accessibility dates, usage rights and whether some form of quality control, like peer review, has been carried out. Some platforms, repositories, registries or funders enforce or encourage the supplementation of records with ample additional information according to the

18 https://www.wihoforschung.de/de/oauni-2182.php

OAI protocol for metadata harvesting[19] to facilitate working with the material. Still, most of the time no or only partial information is available. Moreover, the information at hand is presented in a variety of ways, devoid of universal standards, which could enable a straightforward automated processing of metadata.

This implies that in many cases OA information has to be inferred by other means. In these cases, most studies which have been previously conducted withdraw to performing at least part of the data gathering and classification procedure manually. Because these manual tasks are time consuming, some restrictions are necessary. Often, only a specific subset of all publication output of interest is considered, like a randomly drawn sample of all records or publications from selected years, journals, institutions, countries, publishers or sources (e.g. Taubert 2019). It is particularly difficult to identify Delayed OA, or Moving-Wall OA and to distinguish it from Hybrid OA records. One way is to manually create lists of Full OA, Hybrid OA and Delayed OA journals by using publisher information and manual checks on the publishers' websites. This was done for example in (Laakso & Björk 2016), who limited their analysis to the five largest scholarly publishers Elsevier, Springer, Wiley-Blackwell, Taylor & Francis, and Sage.

However, for large scale analyses a manual categorization is usually not feasible. Therefore, we propose an automated mapping procedure of scholarly publication output to their corresponding categories. Our approach is based upon existing indexes, tools, and harvesting engines and thus, it necessarily inherits the limitations that the mentioned instruments have, especially in terms of accuracy and coverage. For this reason, the first steps in the currently running OAUNI project, that the authors are involved in, are to conduct extensive analytical studies on the most prevalently used resources, namely the Web of Science[20] database of scientific publications and the Unpaywall[21] evidence source for open availability. Despite the limitations, from our point of view the presented method is currently the best compromise between precision and practicability of the classification.

*Full OA publications* are published in cover-to-cover openly accessible journals, allowing immediate access at the time of publication. These journals can be identified by checking them against the ISSN-GOLD-OA 3.0 list provided by Bruns et al. (2019)[22]. The list includes journals from the DOAJ[23], the Directory of OA scholarly resources (ROAD)[24], PubMed Central (PMC)[25], and the Open APC initiative[26] and provides ISSN and ISSN-L information for matching. The DOAJ further lists the year of the transition to open availability, information that the ISSN-GOLD-OA 3.0 list does not provide. The DOAJ is an internationally established evidence source for fully OA peer-reviewed journals with quality assured entries. Exhaustive metadata on the single journals, like the year of transition to OA and the business model are included. The data are openly available and can be obtained using

19 https://www.openarchives.org/OAI/openarchivesprotocol.html

20 https://clarivate.com/products/web-of-science/

21 https://unpaywall.org

22 https://pub.uni-bielefeld.de/record/2934907.

23 Directory of Open Access Journals, https://doaj.org.

24 http://road.issn.org.

25 https://www.ncbi.nlm.nih.gov/pubmed.

26 https://www.intact-project.org/openapc.

the OAI protocol for metadata harvesting.[27] Matching can be performed using the ISSN, taking into account the year of OA-transition.

Articles in subscription-based journals, which are made openly accessible in a formal communication channel immediately at the time of publication, were classified as *Hybrid OA*. For Crossref indexed records, license information provided in form of license URLs can be used to identify Hybrid OA content.[28] Crossref contains extensive metadata including information on the work, authors, journal, publisher funder and – if available – licenses. The metadata can be accessed via an API. Extensive filtering options are offered. The matching can be carried out using the Crossref-doi of the record, the ISSN of the journal and publisher information.

Identifying *Delayed OA* is a challenging task, since it requires information on the respective publishing policies. We want to suggest three approaches here, all of which classify scientific articles on the journal level. The first approach is to use the license node ‘delay-in-days‘ contained in Crossref metadata. It describes the time difference between the earliest publication date and the license statement start date. However, because the Crossref API only allows for querying for delays smaller than a given number of days, an automated search for Delayed OA articles or journals is only possible for subscribers who have access to the whole Crossref metadata dump. A second option is to use Unpaywall information on the OA status of all articles published in a given journal. If the proportion of openly available articles from older issues is close to one, but the journal is not registered in the ISSN-GOLD-OA 3.0 list or the OA share among publications from the recent year is significantly lower, this might be due to a Delayed OA policy. Therefore, comparing the OA proportion of older articles with a registration in the ISSN-GOLD-OA 3.0 list, or with the OA share of the recent year can be used as indicator for a classification of a given journal as Delayed OA. Another possibility is to consult the PMC journal list[29] containing information on the duration until free access. The matching can be performed using the journal title and ISSN. Yet, by design, this collection mainly represents biomedical and life sciences journals. None of the three approaches is ideal, and it might be practical to generate a registry of Delayed OA journals, similarly to the DOAJ or the ISSN-GOLD-OA 3.0 list for fully OA journals.

*Green OA*, that is, publications accessible at places different from the formal communication channel, such as institutional or disciplinary repositories, can be distinguished using harvesting engines like Unpaywall or BASE.[30] Unpaywall (formerly called oaDOI) is an established evidence source for the OA-status of Crossref-indexed articles run by our research[31]. For all objects registered with Crossref, it contains information on the OA-status of the record and – if applicable – the journal, the license, host-type, and more. A matching can be performed using the registered doi, other identifiers, like a PubMed id, or additional information on the title, publication date, journal etc.

---

27 www.openarchives.org/OAI/openarchivesprotocol.html.

28 This approach has been implemented by Jahn (2017) in the Hybrid OA Dashboard.

29 https://www.ncbi.nlm.nih.gov/pmc/journals/.

30 https://unpaywall.org, https://www.base-search.net

31 http://ourresearch.org/

*Table 4: Operationalization Scheme*

| OA Class | Evidence Source | Matching |
|---|---|---|
| Full OA | DOAJ | • ISSN<br>• Year of OA transition<br>• Journal title |
| | ISSN-GOLD-OA 3.0 list | • ISSN<br>• ISSN-L |
| Hybrid OA | Crossref | • Crossref-doi<br>• ISSN<br>• Publisher<br>• Journal title<br>• Publication title<br>• Publication date<br>• Author information |
| Delayed OA | Crossref: license-delay field | • Crossref-doi<br>• ISSN<br>• Publisher<br>• Journal title<br>• Publication title<br>• Publication date<br>• Author information |
| | PMC journal list | • ISSN<br>• Journal title |
| | Unpaywall: Comparison of OA shares amoung recent and older articles published in the same journal | • Crossref-doi<br>• Journal title<br>• Publication title<br>• Publication date<br>• Author information |
| Green OA (for all types, including institutional and disciplinary Green OA as well as preprints and postprints) | Unpaywall | • Crossref-doi<br>• Journal title<br>• Publication title<br>• Publication date<br>• Author information |
| | BASE | • Crossref-doi<br>• Publication title<br>• Journal title<br>• Publication date<br>• Author information<br>• Repository timestamp<br>• Repository identifier |
| | OpenDOAR | • Repository url<br>• OAI-PMH url |
| Non-OA articles | Without evidence | |

Similarly, BASE is an academic search engine run by the University of Bielefeld, which indexes metadata – including information on the OA-status and available licenses – from validated sources[32]. Just like for Unpaywall, the matching of records can be performed using digital identifiers or metadata information like the title, publication date or other tags. The distinction between institutional and disciplinary repositories can be made based upon the BASE data or by matching the repository url provided by Unpaywall with the repository type as specified in the service OpenDOAR. Preprints and Postprints can be distinguished by comparing – if available – the repository timestamp with the date of publication.

The remaining records contain articles in subscription-based journals which are not openly available, articles which are available only on academic social networks, like, for example, Research Gate, or through copyright violations, and all records which are not covered by any of the above categories and they are classified as *Non-OA*.

## 6. Conclusion

In this paper, we identified the need for standards to enable a rigorous investigation and monitoring of the development of OA publishing, suggested a non-normative understanding of OA and developed a classification that helps to distinguish the OA classes that are most important from our point of view. Furthermore, we proposed an operationalisation method based on open evidence sources, which can be used to apply the classification scheme to bibliometric data. We do not expect that all actors in the field will simply agree on the proposed categories but we very much hope that it contributes to an ongoing discussion about how to make OA shares and numbers more comparable.

[32] https://www.base-search.net/about/en/about_sources.php